\documentclass{vgtc}                          % final (conference style)
\graphicspath{{figures/}{pictures/}{images/}{./}} % where to search for the images

\usepackage{times}                     % we use Times as the main font
\usepackage{tabu}                      % only used for the table example
\usepackage{booktabs}                  % only used for the table example
\usepackage{lipsum}                    % used to generate placeholder text
\usepackage{mwe}                       % used to generate placeholder figures

\usepackage{mathptmx}                  % use matching math font

\usepackage{enumitem}
\setlist{nosep}
\usepackage{array}

\usepackage[compact]{titlesec} % Compact section spacing

\usepackage{dirtytalk}

\newcommand{\sys}[1]{\textit{GuidedStats}}
\MakeRobust{\sys}

\usepackage[svgnames, table]{xcolor}

\definecolor{e_orange}{RGB}{248, 229, 208}
\definecolor{e_green}{RGB}{220, 233, 213}
\definecolor{e_blue}{RGB}{166, 196, 229}
\definecolor{e_purple}{RGB}{216, 211, 231}
\definecolor{e_pink}{RGB}{230, 210, 219}

\newcommand{\yuqi}[1]{#1}

\usepackage{quoting}
\quotingsetup{vskip=2pt}

\onlineid{0}

\vgtccategory{Research}

\vgtcinsertpkg

\title{Guided Statistical Workflows with Interactive Explanations and Assumption Checking}

\author{Yuqi Zhang\thanks{e-mail: yz9381@nyu.edu}\\ %
        \scriptsize New York University %
\and Adam Perer\thanks{e-mail: adamperer@cmu.edu}\\ %
     \scriptsize Carnegie Mellon University %
\and Will Epperson\thanks{e-mail: willepp@cmu.edu}\\ %
     \scriptsize Carnegie Mellon University}

\teaser{
  \centering
  \includegraphics[width=0.89\linewidth]{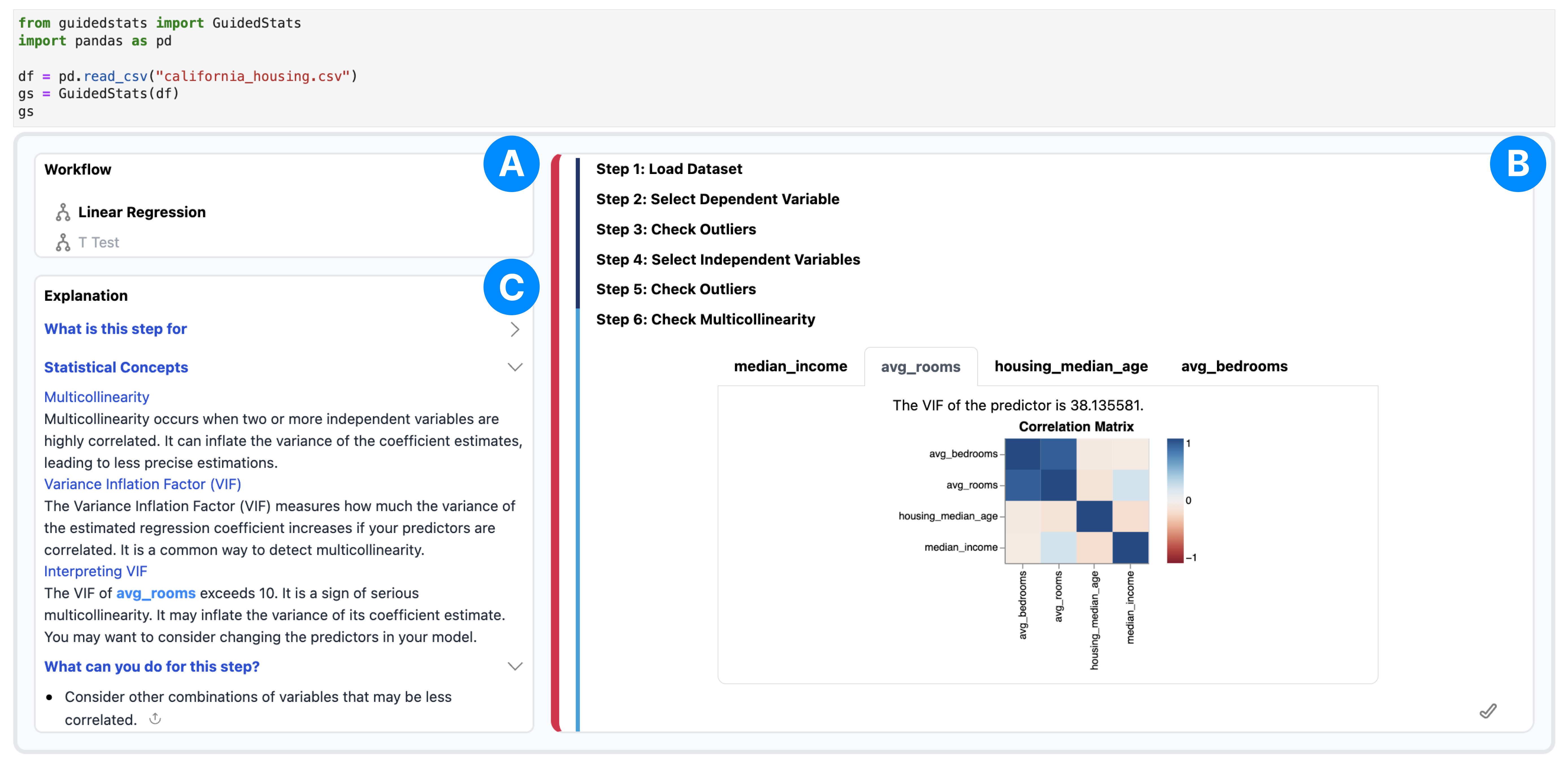}
  \caption{
\sys{} helps users perform statistical analyses with guided workflows. 
(A) A user selects to run the linear regression workflow. 
(B) \sys{} guides the user through the nine steps of this workflow. Each step either solicits user input or helps users verify assumptions with visualizations and tests of the data.
(C) \sys{} provides explanations for the current step in the workflow. The user is currently on an \textit{assumption checking step}; the explanation shows more information about the statistical concepts in the current step.
}
  \label{fig:overview}
}

\abstract{
Statistical practices such as building regression models or running hypothesis tests rely on following rigorous procedures of steps and verifying assumptions on data to produce valid results. 
However, common statistical tools do not verify users' decision choices and provide low-level statistical functions without instructions on the whole analysis practice.
Users can easily misuse analysis methods, potentially decreasing the validity of results.
To address this problem, we introduce \sys{}, an interactive interface within computational notebooks that encapsulates guidance, models, visualization, and exportable results into interactive workflows. 
It breaks down typical analysis processes, such as linear regression and two-sample T-tests, into interactive steps supplemented with automatic visualizations and explanations for step-wise evaluation.
Users can iterate on input choices to refine their models, while recommended actions and exports allow the user to continue their analysis in code.
Case studies show how \sys{} offers valuable instructions for conducting fluid statistical analyses while finding possible assumption violations in the underlying data, supporting flexible and accurate statistical analyses.
}

\keywords{Data science tools, computational notebooks, analytical guidance}

\begin{document}


\firstsection{Introduction}

\maketitle
Statistical analysis is widely used to guide decision making and make sense of experimental results.
However, producing \textit{valid} statistical estimates requires following particular statistical procedures with specific data assumptions~\cite{Agresti2009StatisticalMF,stock2020econ}.
In practice, even basic analysis methods like null hypothesis testing or linear regression models are misused, leading to potentially incorrect analysis results~\cite{
Ernst2017RegressionAI,Hoekstra2012AreAO,thiese2015misuse,Vornhagen2020StatisticalST}. 

One of the ways statistical methods are misused is when assumptions are not properly verified~\cite{thiese2015misuse}.
All statistical models rely on assumptions about the input variables.
If these assumptions are heavily violated, the statistical estimates may be biased and inconsistent, potentially leading to incorrect inferences and misguided decision making~\cite{Ernst2017RegressionAI,Osbourne2002FourAO}.
For example, during a two-sample T-test, analysts may neglect to check if the variances of the two samples are equal. 
The t statistic is calculated differently depending on whether variances are equal or not, and an incorrect t statistic might lead to a false rejection of the null hypothesis.

% For example, if the means for two paired sample data are compared with student's T-test, the statistical results are invalid to support their conclusion for experiments~\cite{thiese2015misuse}.

% Another reason leading to the difficulty in applying statistical methods is that the resources for \textit{how} to do the analysis are separate from \textit{where} the analysis takes place.
Correctly using statistical methods is difficult in part because the resources for \textit{how} to do the analysis are separate from \textit{where} the analysis takes place.
Numerous textbooks and frameworks exist describing the assumptions that must be met when using statistical tools like hypothesis tests or linear regressions~\cite{Hahn1993AssumptionsFS,Jun2021HypothesisFE}.
However, many analysts perform their analyses in coding environments like computational notebooks, where they can manipulate, visualize, and run statistics on their data using various packages.
These code-based analysis environments are preferred over GUI-based tools since programming affords more flexibility in the analysis~\cite{perkel_2018, Rule2018ExplorationAE}.
Despite their flexibility, these statistical analysis packages can be easily misused \yuqi{as they offer little assistance in ensuring correct manipulations.
% \textbf{guidance} during the analysis.
Providing guidance can bridge the knowledge gap necessary for users to make analysis progress and thus perform correct statistical practices~\cite{Ceneda2024AHA,Ceneda2017CharacterizingGI}.
% In this scenario, it takes the form of explanations and instructions to help users verify assumptions and transform their data if those assumptions are violated.
}

In this paper, we introduce \sys{}, an interactive system within computational notebooks to guide analysts through statistical workflows.
\sys{} contains different workflows for common statistical procedures that walk a user through steps to specify inputs for their analysis and check necessary assumptions.
% \sys{} begins its guidance with a statistical analysis workflow.
% The workflow maintains a sequence of steps that proceeds either linearly or iteratively through the analysis.
% The workflow verifies assumptions of models with statistical tests and visualization.
\sys{} offers guidance in two primary ways: by walking a user through the proper steps in a workflow and explaining how to interpret each step.
% Each step has explanation to help users understand the objectives of steps and interpret statistical results.
Step explanations include the step's objective, statistical concepts, and how to interpret the results.
\yuqi{
After checking assumptions, \sys{} offers suggestions based on the results. 
Each suggestion includes actions that can populate later inputs or export code to transform the data to meet the assumptions.}
% \yuqi{Suggestions for the assumption results are offered with recommended actions to take if the assumptions are violated.
% Actions can populate inputs in later input steps or export code to transform the data to try to meet the assumption.}

% \TODO{add another paragraph with details about the steps, kinds of workflows, and why explanations are interesting}
As a Jupyter Notebook Widget, \sys{} allows fluid exchanges between the user interface and the coding environment.
\sys{} takes in a dataset for analysis; once the statistical workflow is complete, users can export the results back to code.
During the analysis, users can explore or change the dataset used by \sys{} with code.

We present two use cases where \sys{} helps users find assumption violations and possible actions for the assumption checking results on linear regression and two sample T-test workflows.
Our code is open source and available for use\yuqi{\footnote{\url{https://github.com/cmudig/GuidedStats}}}.

In summary, our paper makes the following contributions.
\begin{enumerate}
    \item A framework for guided statistical analysis with automatic assumption checking and explanations. Assumption checks help users ensure their analysis meets expected assumptions, while explanations help users understand the concepts and interpret the results at each step. Assumption violations can be corrected through recommended actions on the data.
    \item \sys{}, a computational notebook extension that incorporates these guidance elements into an interactive system for guided statistical workflows and supports hand-off between the interface and code.
\end{enumerate}

\section{Related Work}
% \sys{} builds on prior research on software designed for guided statistical analysis and takes the form of Widget in Jupyter Notebook.

% \subsection{Guided statistical analysis}
Commonly used statistical analysis tools such as SPSS, SAS, Stata, and R, provide UIs for statistical analysis, \yuqi{but they often lack sufficient explanations and documentation for users to effectively perform and understand statistical procedures~\cite{Hodges2022ResearcherDO}.}
% ambiguous explanations and documentation.
This lack of support limits analysts in understanding the underlying concepts behind the methods they use~\cite{Wild1999StatisticalTI}.
Furthermore, these tools may not verify the assumptions of the statistical models the analysts choose, which leads to unreliable statistical estimates.
% Therefore, it is important to incorporate sufficient guidance and assumption checking into statistical tools to address potential  threats to validity.

Researchers have developed statistical tools that offer step-by-step guidance and verify assumptions during analysis.
\yuqi{\textit{StatHand} is a web-based application designed to help students identify appropriate statistical tests and provide guidance on using statistical tools and explaining statistical concepts~\cite{Allen2016IntroducingSA}.
However, \textit{StatHand} can not perform statistical analysis within its interface or offer guidance on decisions made during the analysis.
\textit{Statsplorer} allows users to perform statistical tests within its web-based application, explaining statistical concepts and verifying necessary test assumptions~\cite{W2015Statsplorer}.
While these tools can explain statistical concepts and interpret model results, they do not guide decision-making during analyses, such as managing assumption checking results. 
In contrast, \sys{} provides guidance specific to the current dataset and recommends actionable steps for handling assumption violations.}

\yuqi{
\textit{Tea} and \textit{Tisane} offer high-level declarative languages for authoring statistical analyses~\cite{Jun2019TeaAH,Jun2022TisaneAS}. 
They help users specify hypotheses, study designs, and variable relationships, automatically verifying assumptions and inferring valid models.
While \textit{Tisane} provides suggestions for assumption violations, it focuses on explaining concepts in assumptions, requiring manual implementation and risking misuse.
In contrast, \sys{} guides the entire statistical analysis process by explaining each step’s objectives and suggesting actions based on assumption verification results.
\sys{} also supports iterative model development, allowing users to revise decisions, tightening the feedback loop of model refinement.
}

% Programming languages offer high flexibility for data analysis due to their expressiveness, while graphical interfaces provide efficient and easy-to-use parameterization of data analysis modules\cite{Alspaugh2019FutzingAM}. 
% To fully avail their advantages, users often switch between coding and graphical tools, which can be tedious because of transferring code and data from one another\cite{Chattopadhyay2020WhatsWW}, causing usage friction\cite{Kandel2012EnterpriseDA}. 
% It can shift the focus from analysis to downstream implementation tasks and force the trade-off between code and interface.
% Various tools(e.g. \cite{Wu2020B2,Kery2020mageFM,Epperson2023DeadOA}) have attempted to moderate the trade-off by enabling bi-directional communication, that is receiving input from the notebook while updating the notebook content\cite{Wang2023SuperNOVADS}.
% \sys{} adopts bi-directional communication, allowing the import of processed datasets and the export of investigatable outputs within the analysis workflow.
Current interactive statistical tools exist apart from where many analyses actually take place: computational notebooks.
Computational notebooks integrate text, code, and visualization into a single document, supporting flexible and iterative analyses~\cite{perkel_2018,Rule2018ExplorationAE}.
% Among these, Jupyter Notebooks are particularly popular with data scientists~\cite{kaggle2022}.
% Jupyter includes Widgets to embed interactive GUIs into notebook.
% It synchronizes information between Python and JavaScript and thus enables to transfer state variables of GUIs to the coding environment~\cite{ipywidgets2023}.
% This connection between GUI and coding environment makes Jupyter Widgets both easy to use and flexible.
Prior systems extend computational notebooks to better support revisiting history in analysis and data profiling~\cite{Epperson2023DeadOA,Kery2020mageFM,Wu2020B2}.
Likewise, \sys{} is a computational notebook extension to support easily jumping between GUI and code.
A user provides an initial dataset to \sys{}, which can be later updated if the data needs transformation after an assumption violation is discovered.
At the end of an analysis, the results can be exported back to code.
% It shares dataset with the notebook, accepting both data manipulation in the coding environment along with performing statistical analysis on the data in GUI.
% It can also export statistical results to notebook in different formats to help document the analysis process.
% \TODO{talk somewhere about what is jupyter, and that it is the tool of choice for computational notebook for data science}

% \sys{}, in contrast, offers comprehensive guidance on statistical concepts, interpretation, purposes of steps,and suggestions. 
% It interprets statistical results and supports reasoning about  actions taken by both users and \sys{} during statistical practice. 
% It verifies the decision choices by testing assumptions and allows iterative development of models, leveraging the easy-to-use quality of interfaces.
% Its framework enables the extension of more statistical practice.

\section{\sys{}: step-by-step statistical workflows}
\sys{} is a computational notebook extension that has statistical \textbf{workflows} comprised of \textbf{steps} and \textbf{explanations}.

\subsection{Statistical workflows}
To start using \sys{}, a user selects which statistical \textbf{workflow} they want to use.
Each workflow represents a statistical analysis and guides users through the steps in the workflow.
\sys{} currently contains two workflows: Linear Regression and T-Tests, but can be extended in the future to other \yuqi{statistical workflows}. 
\yuqi{
% Each statistical workflow consists of sequential steps where each step’s required intermediate results are calculated in the preceding steps.
Workflow steps are tasks such as choosing input variables and checking model assumptions.
Other statistical methods can be formulated as workflows in \sys{} by specifying the necessary steps to gather inputs and assumptions.
}

Workflows also support the iterative refinement of decision choices.
Every time a previous step is edited, the following steps are rerun with the new parameters, enabling quick iteration when testing different decision choices.
Additionally, a workflow manages imports and exports for the analysis.
As input, the workflow ingests the initial dataset.
Once the analysis is complete, a user can export the created model and the report of the results.
During the analysis, workflows can export code for each ongoing step and the dataset.
If a dataset is transformed in the notebook, it can be re-imported back to the workflow, which will re-run the analysis up to the current step.

% To facilitate iterative development, users can make different decision choices for steps that have already been completed.
% workflow receives the new input and reconfigures all subsequent steps keep other decision choices constant, enabling quick evaluation of different hyperparameter choices.

% Additionally, workflow handles imports and exports.
% It receives a dataset through code and passes it to the designated step for Dataset Loading.
% If a dataset is imported during the analysis, workflow will rerun the analysis to maintain alignment between \texttt{workflowInfo} and the dataset.
% On the other hand, whenever an export is requested - whether for the dataset, the model, the model interpretation report, or the code of the ongoing \textbf{workflow} - workflow will refers to the step responsible for the export and export items into code below the widget cell.

% \TODO{this is too much about the implementation right now. Implementation part about importing workflows should only be a couple sentences -- I want to know more about the conceptual model of the workflow.  E.g. conceptually a workflow is comprised of steps that you have to go through. Is there anything else interesting or novel? We have two example workflows now -- can every single statistical process be thought of as a workflow, what is the limit?}

\subsection{Steps in a workflow}
\label{sec: steps}

\begin{table*}[h]
    \setlength{\extrarowheight}{1.2pt} % Adds extra space at the top of table rows
    \centering
    \begin{tabular}{p{0.12\linewidth}p{0.22\linewidth}p{0.58\linewidth}}
        Step Type & Purpose & Explanation Components\\
        \hline
        \textbf{User Input Step} & Require user inputs from the interface to specify their decisions 
        &
        \vspace{-15pt} % remove weird spacing 
        \begin{minipage}[t]{\linewidth}
        \begin{enumerate}[nosep,leftmargin=*] 
            \item \textit{Objective:} purpose of the step
            \item \textit{Statistical Concepts \& Interpretation:} explanations of important statistical concepts used in the step
        \end{enumerate} 
        \end{minipage}
        \vspace{1pt}
        \\
        \hline
        \textbf{Result Presentation Step} &  Presents the outcomes of the statistical analysis
        & 
        \vspace{-15pt} % remove weird spacing 
        \begin{minipage}[t]{\linewidth}
        \begin{enumerate}[nosep, leftmargin=*]
            \item \textit{Objective:} purpose of the step
            \item \textit{Statistical Concepts \& Interpretation:} interpretation of the modeling results
        \end{enumerate} 
        \end{minipage}
        \vspace{1pt}
        \\
        \hline
        \textbf{Assumption Checking Step} & Verify the assumptions of models to ensure
    the validity of statistical estimates  
        & 
        \vspace{-15pt} % remove weird spacing 
        \begin{minipage}[t]{\linewidth}
        \begin{enumerate}[nosep, leftmargin=*]
            \item \textit{Objective:} purpose of the step
            \item \textit{Statistical Concepts \& Interpretation:} interpretation of the results of tests used to verify the assumptions
            \item \textit{Suggested Actions:} possible actions if the assumption is violated
        \end{enumerate} 
        \end{minipage}
        \vspace{1pt}
        \\
    \end{tabular}
    \caption{The types, purpose, and explanation components of steps. See \autoref{sec: explanation} for the example of each.}
    \label{tab:steps}
\end{table*}

In \sys{}, workflows are comprised of individual \textbf{steps}. 
Each step represents a sub-task in the analysis.
For instance, as shown in \autoref{fig:overview}, the linear regression workflow includes steps for identifying independent variables (Step 4) and verifying the assumption of outliers (Step 3 and 6).
The steps, assumptions, and visualizations in the current \sys{} workflows are adapted from statistics textbooks~\cite{Agresti2009StatisticalMF,hanck2021introduction,stock2020econ} and further refined using examples from online analysis packages like scikit-learn~\cite{scikit-learn} or Vega-Lite~\cite{vega-lite-examples}.

There are three kinds of workflow steps in \sys{}: User Inputs Steps, Result Presentation Steps, and Assumption Checking Steps (\autoref{tab:steps}).
User Input Steps enable users to set parameters in the analysis, such as selecting independent variables.
Result Presentation Steps are always the last step in an analysis and present the results of a workflow.
% It occurs only once at the end of the workflow and displays tables and visualizations of model parameters and evaluation.
% The decisions include selecting variables, setting hyperparameters, or choosing models.
Assumption Checking Steps verify the assumptions of a workflow.
For example, the linear regression workflow includes checks for outliers and multicollinearity.
% \yuqi{sometimes also for homoscedasticity and strict linearity}.

% User Input Steps primarily affect accuracy through different inputs of decision choices.
Assumption Checking Steps are important for ensuring valid statistical models and informing user actions.
For instance, if the data fails the normality assumption in a two-sample T-test, users might transform the data or choose a non-parametric test instead.
Assumption Checking Steps remind users of important assumptions, which they can choose to ignore depending on the analysis context.
% However, users may sometimes ignore minor violations in specific scenarios to preserve the original data.
For example, with a large sample size, the robustness of the T-test allows users to overlook non-normal distributions of two groups. 
Violations in Assumption Checking Steps are accompanied by Suggested Actions that users can take to potentially fix the violation.
% Thus, instead of \yuqi{prescribing} actions for handling violations, Assumption Checking Steps provide \textbf{Suggested Actions} in the explanation to recommend common actions for dealing with the violation.

\subsection{Step Explanations}
\label{sec: explanation}

\textbf{Explanations} are offered for each step in \sys{} and have three potential components: the objective of the step, the statistical concepts \& interpretation, and suggested actions.
\autoref{tab:steps} details the components of the explanation for each type of step.

All steps have an objective explanation and a statistical concept explanation.
The objective of the step clarifies why the step is included in the workflow and its importance to the overall statistical analysis.
For instance, the step for stating the hypothesis in the two sample T-test has the objective ``Formulate the null and alternative hypotheses for the T-test and define the Type I Error (alpha)''.

The statistical concepts and interpretation help users interpret the results in each step.
This type of explanation is included in each step but is particularly important for the result presentation step.
For example, in the result presentation for a linear regression, the explanation shows how to interpret the effect of the coefficients and the metrics for measuring the fitness of model, such as $R^2$.

Suggested actions are provided only for Assumption Checking Steps.
% \yuqi{They are derived from the same resources as assumptions, using templates as Epperson et al.~\cite{Epperson2023DeadOA} to offer suggestions based on the assumption verification results.}
\yuqi{Like the steps themselves, these actions are derived from standard statistical textbooks~\cite{Agresti2009StatisticalMF,hanck2021introduction,stock2020econ}.
% They are presented as code templates adaptable to the assumption verification results, as Epperson et al.~\cite{Epperson2023DeadOA}}.
Actions take two forms: setting up the inputs for parameters in future steps or generating code to fix an assumption violation inspired by prior approaches for exportable analysis snippets \cite{Epperson2023DeadOA}.}
For example, when checking the linear regression assumption of multicollinearity, a suggestion is presented as "consider other combinations of variables that are less correlated".
Clicking on that action will produce a code cell to drop the variables with the highest correlation.
\section{Use cases}
\label{sec: case study}
In this section, we describe two example statistical analyses using \sys{} with real-world datasets: a multivariate linear regression and a two-sample T-test.
\yuqi{In these examples, the analysts are novices in statistics but aim to perform reliable statistical analyses for their reports.}
These analyses illustrate how \yuqi{they can} benefit from the design and guidance of \sys{} during the analyses. 

\subsection{Linear Regression}
In our first use case, we discuss an analyst who wants to understand how housing characteristics such as housing age and neighborhood income affect house value.
For their analysis, they use the California Housing Prices Dataset, which contains 20,640 observations of the median house value for different blocks of houses in California~\cite{Pace1997SparseSA}.
They want to use a linear regression model to help them estimate the impact of different features on the median house value, so turn to \sys{} to create the model.

\begin{figure}
    \centering
    \includegraphics[width=0.86\columnwidth]{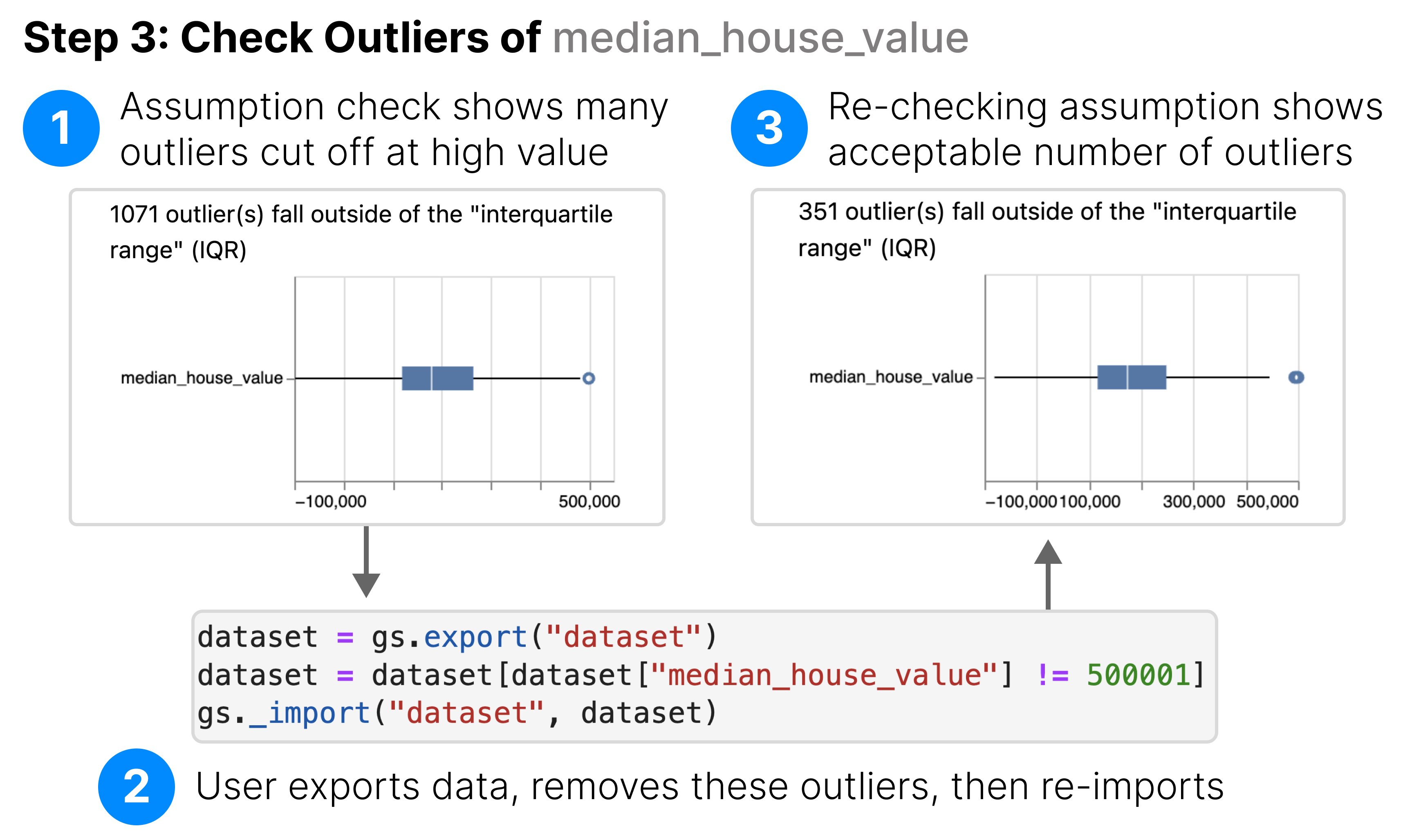}
    \caption{\sys{} supports an interactive loop of assumption checking, editing the data, then re-verifying assumptions.}
    \label{fig:edit data}
\end{figure}

First, the analyst imports the dataset and selects the Linear Regression Workflow. 
The workflow contains 9 steps, including 3 assumption checking steps to check the core assumptions of linear regression: no large outliers in the independent variables or dependent variables, and no perfect multicollinearity~\cite{stock2020econ}.
After loading the data, the second step is variable selection, where the analyst selects \texttt{median\_house\_value} as the dependent variable from the displayed list of columns.
\sys{} then moves to Step 3 to help the analyst verify the first assumption: that no large outliers exist in the target variable.
This step plots the distribution of \texttt{median\_house\_value} and indicates that 1,071 outlier points lie outside the interquartile range (a common method for checking outliers).
The explanation block of this step suggests actions to deal with these outliers, such as removing the outliers or applying data transformation to the column, like a log transform.
The analyst follows the suggestion to investigate the column first and clicks the action button of this suggestion.
\sys{} exports code for descriptive analysis on \texttt{median\_house\_value} to a new cell below.
The output reveals that the most frequent value is 500,001, where \texttt{median\_house\_value} was manually clipped. %according to the dataset description.
The analyst then removes rows with \texttt{median\_house\_value = 500,001}, and re-imports the transformed dataset to \sys{}~(\autoref{fig:edit data}).

Next, in Step 4 (Variable Selection Step), the analyst selects \texttt{median\_income}, \texttt{total\_rooms}, \texttt{total\_bedrooms}, \texttt{housing\_median\_age} as independent variables to see how they impact the median housing price.
In the next step, \sys{} once again helps the analyst check the assumption that no large outliers exist in the independent variables either.
Inspecting the output from this step shows that some selected variables, \texttt{total\_rooms} and \texttt{total\_bedrooms}, have a large amount of outliers and thus might jeopardize the validity of the regression model if used.
After reviewing the dataset description, the analyst decides to create more meaningful variables \texttt{avg\_rooms} and \texttt{avg\_bedrooms} as the average of the rooms and bedrooms for all houses in the block and use these as independent variables. % instead of total number of rooms in the block
% \texttt{median\_income}, \texttt{avg\_rooms}, \texttt{avg\_bedrooms}, \texttt{housing\_median\_age} as independent variables.
% For instance, \texttt{total\_rooms} represents the number of rooms in the whole block.
% It is not a characteristic of the house itself.
% Therefore, like the original paper~\cite{Pace1997SparseSA}, the analyst creates \texttt{avg\_rooms}, the average room of the population.

After ensuring no large outliers, \sys{} moves to another assumption checking step to check for multicollinearity by calculating Variance Inflation Factor (VIF) of each independent variable. 
This metric measures the correlation among the independent variables and its effect on accuracy.
\sys{} shows that \texttt{avg\_rooms} and \texttt{avg\_bedrooms} have high VIFs (38.14 and 30.57), indicating significant multicollinearity, though they may still be valid depending on the context (\autoref{fig:overview}).

The analyst notes this issue and proceeds to the next two user input steps: data splitting and model hyperparameter selection. 
In the model specification step, they choose a simple linear regression.
% The analyst then moves on to the last step in the workflow for model evaluation.
In the final model evaluation step, \sys{} displays the coefficients and goodness-of-fit measures, such as $R^2$. 
The $R^2$ value, at 49.4\%, indicates that approximately half of the variance in \texttt{median\_house\_value} is explained by selected independent variables.
The positive coefficient for \texttt{avg\_bedrooms} suggests that the median house price increases as the average number of bedrooms increases. 
Conversely, a negative coefficient for \texttt{avg\_rooms} indicates as the average number of rooms increases, the median house price decreases. 
This somewhat contradictory result is likely due to the multicollinearity observed earlier, which may undermine the model's interpretive power - a major concern. 
This highlights the need for further refinement of the model.

The analyst continues iterating on different combinations of independent variables in Step 4 (Variable Selection Step) while keeping other decision choices unchanged.
They finalize a model with independent variables \texttt{median\_income},  \texttt{avg\_rooms}, \texttt{housing\_median\_age} and \texttt{households}.
The model produced with these features has a slightly lower $R^2$ (0.47 compared to 0.49) than before.
However, the analyst would rather use this newer model since there is no multicollinearity between the features.
% These independent variables show no multicollinearity issue in checking this assumption despite slightly lower $R^2$(0.4717 compared to 0.4936) shown in Step 9, which is less concerning than the multicollinearity.

\subsection{Two Sample T-Test}
In our second use case, we describe an analysis of car data to test if miles per gallon (mpg) of cars differs between regions.
This analysis uses the cars dataset, which contains 398 observations with 8 attributes about cars~\cite{misc_auto_mpg_9}.
Our analyst tests the null hypothesis that the cars have no difference in \texttt{mpg} between the US and Europe.

The analyst first imports the dataset into \sys{} and selects the two sample T-test workflow.
This workflow has 9 steps, with 4 of them to check the 3 core assumptions of a T-test.
The workflow begins by asking the analyst for the variable where they select \texttt{mpg}.
Next, \sys{} guides the analyst to check the first assumption of a T-test: lack of outliers.
Here, \sys{} displays the same distribution overview used in the prior use case, showing only 1 outlier in \texttt{mpg}.
Finding this acceptable, the analyst moves on to the next step, which is to select the two groups to compare in the T-test.
They first select the column \texttt{origin}, then the values for US and Europe as the groups to compare.
% In Step 4(\texttt{Variable Selection Step}) to find group variable to split \texttt{mpg}, she notices that the column \texttt{origin} is not listed in the selection panel.
% \sys{} in the \texttt{Explanation} block inspires her that the column \texttt{origin} may not be explicitly implied as a categorical variable in the dataframe.
% As a result, she exports the dataset, sets the type of \texttt{origin} as 'category', and imports the dataset back.
% Then she can select the first two groups of \texttt{origin}.
\begin{figure}
    \centering
    \includegraphics[width=0.86\columnwidth]{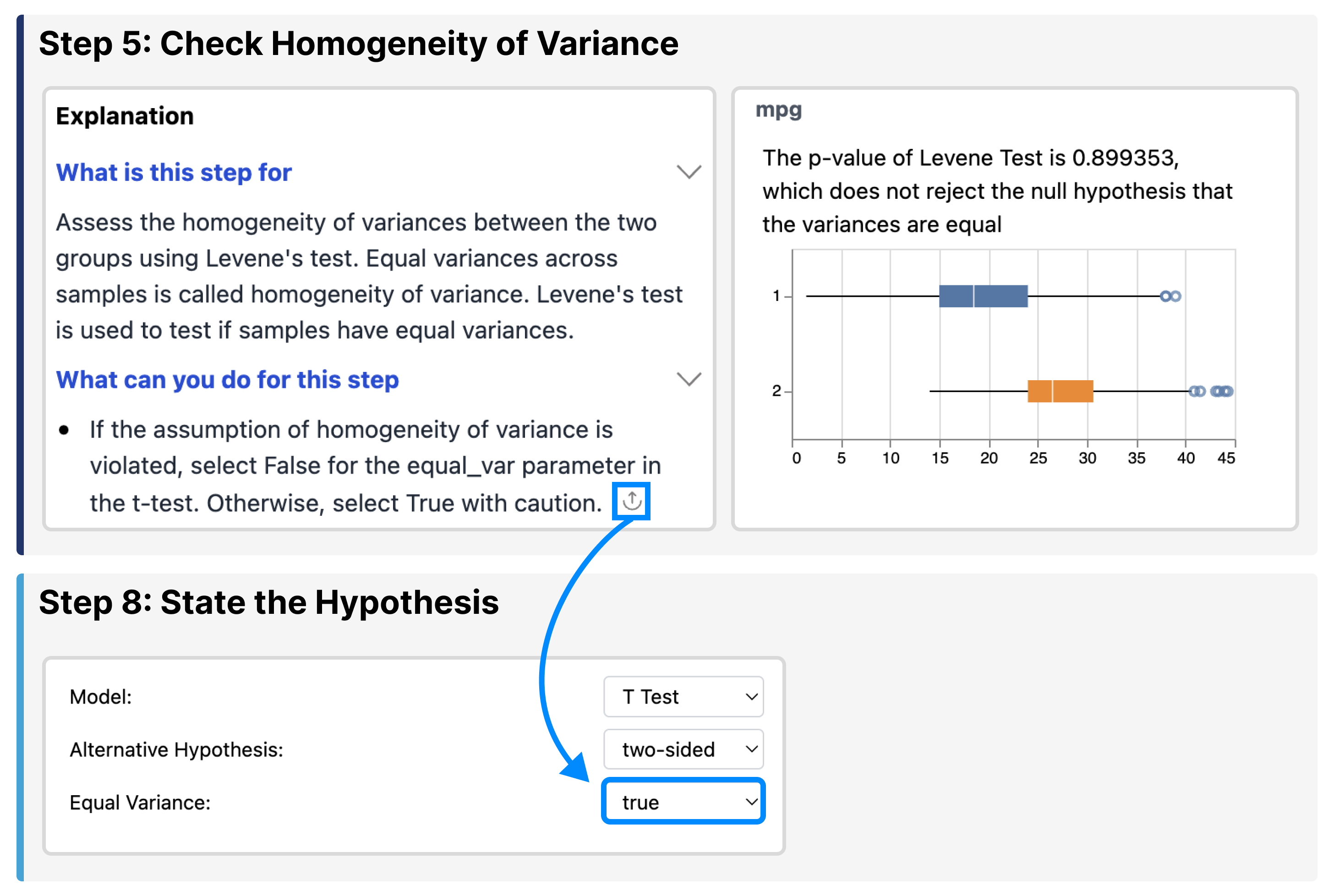}
    \caption{
    For assumption checking steps, like the homogeneity of variance in the T-test workflow, \sys{} recommends potential actions based on the results of assumption checks.
    In this example, the check suggests the variances are equal and the user can select the \textbf{action} to pre-set this parameter to True in the later model specification step.
    % Checking homogeneity of variance.  {\large \textcircled{\small 1}} \sys{} displays the p-value from Levene’s test along with a boxplot showing the distributions of the two groups. 
    % {\large \textcircled{\small 2}} In the Explanation panel, \sys{} recommends setting the "equal\_var" parameter of the T-test based on the test results. An action button is provided to apply this setting in advance to the UI. 
    % {\large \textcircled{\small 3}}  After clicking the action button, the "equal\_var" parameter for the T-test is pre-set to True.
    }
    \label{fig:ttest_group_variance}
\end{figure}

\sys{} continues to the next step to check the next assumption for T-Tests: the homogeneity of variance between groups.
Since the analyst is new to T-Tests, they are unsure what this assumption is checking and turn to the explanation block to learn more. 
% \sys{} helps the analyst with the interpretation of the new statistical concept with \texttt{Explanation}.
\sys{} explains that homogeneity of variance assesses whether the comparing groups have similar variances and can be evaluated using Levene's test~\cite{Gastwirth2009TheIO} (\autoref{fig:ttest_group_variance} top).
\sys{} plots the distributions of the two groups along with Levene's test result. 
The p-value of Levene's test is 0.90 and doesn't support rejecting that variances are equal.
As suggested, the analyst clicks on its action button to pre-select the parameter for "equal variance" of the T-test as True in the later model specification step~(\autoref{fig:ttest_group_variance} bottom).

In Steps 6 and 7 (Assumption Checking Steps), \sys{} helps the analyst check the final assumption of T-test on whether the two groups are normally distributed. 
The analyst initially observes that the distributions of the two groups are right-skewed in the density plot and is concerned that this assumption may be violated.
However, \sys{} includes a suggestion: ``If the sample size is large enough (greater than 30 as a rule of thumb), the T-test is robust to violations of normality''.
When the analyst clicks on the action button for this suggestion, \sys{} shows them a notice stating that the sizes of two groups are both sufficiently large to be treated as normal.
Therefore, the analyst bypasses the assumption and proceeds to Step 8 (Model Step).
Here they select ``two-sided'' as the alternative hypothesis and retain the pre-selected ``equal variance'' parameter set to True from the earlier action.
In the last step for evaluation, \sys{} plots the mean difference of two groups and shows a t-statistic of -8.9147 with p-value close to zero. 
Consequently, the analyst rejects the null hypothesis and concludes that there is no difference in \texttt{mpg} between cars from the US and Europe.
% \TODO{Conclusion sentence about how \sys{} helped this analysis, somethign like -- In this analysis, \sys{} helped the example analyst check assumptions about their data and notice X,Y that would have been difficult to see without the help of \sys{}.}

In these analyses, \sys{} helps the analysts identify and deal with outliers and multicollinearity in the linear regression model and verifies the assumptions of equal variance and normality in the two-sample T-test. 
The interpretation and suggested actions, such as interpreting high VIF scores and presetting the "equal variance" parameter, would have been difficult and time-consuming to manage without the support of \sys{}. 
\section{Discussion and Future Work}

In statistical analysis, verifying assumptions is important to ensure the validity and reliability of results.
\sys{} provides a framework incorporating assumption verification with actionable explanations in the analysis workflow.
It ensures that the assumptions of statistical models are well-supported by test results, visualizations, and explanations with suggested actions.

However, some assumptions can not be tested numerically as they extend beyond the dataset itself.
For example, linear regression assumes \yuqi{variables are independently and identically distributed} (i.i.d.)~\cite{stock2020econ}.
To verify this, users must \yuqi{know the data collection process prior to their analyses.}
% \sys{} verifies assumptions mostly with only the dataset and thus does not cover the assumptions that are not numerically testable.
While \sys{} could suggest examining the randomness of data collection or export exploratory data analysis code, \yuqi{these suggestions are not actionable from the system as they are not derived from the dataset or research design}.
Adding such steps could increase the ambiguity of statistical analyses and introduce new statistical concepts without clear interpretation or actionable guidance. 
\yuqi{\sys{} focuses on the "must-have" assumptions for each workflow. 
Other assumptions that are not numerically testable or ``optional'' for statistical validity  (such as the normality of error terms for statistical derivation) are not incorporated. 
Future work might explore how to also incorporate these other less-strict versions of assumptions.}

% Therefore, in \sys{} users have to verify numerically untestable assumptions themselves, which may still leave a threat to validity.
% Future work could expand to supporting assumptions that are difficult to verify by numerical calculation, which is the basis of current statistical tools.

% \sys{} is a computational notebook extension that leverages the user interface and coding environment to assist in performing statistical analyses.
% It provides guidance and executes the statistical analysis in its interface, extended with the coding environment by sharing state variables.
% However, the interface is one of the ways to guide analyses.
% Other forms may include monitoring the coding environment and noticing the users when there's an assumption for checking, or in a Q\&A manner where users specify their decision choices in natural language and the system returns the statistical results by step.
% Future work could explore the forms of providing guidance and evaluate their best usage scenario.

As a computational notebook extension, \sys{} \yuqi{facilitates transitioning between code-based analysis and a user interface}.
% However, its guidance and explanations are currently constrained by the interface.
Future work could enhance the pairing of code-based and GUI-based analysis such as monitoring users' code and alerting them to potential statistical issues.
Additionally, other interfaces such as natural language chat-based tools could allow users to ask questions about their workflow or data, \yuqi{further augmenting statistical programming}.

\acknowledgments{
Many thanks to the anonymous reviewers for their insightful suggestions. We also thank Jiayuan Song, Richard Lin, Zhuoqing Li, and Zhuo Zeng for their valuable feedback and inspiration during the development of the system.}

\bibliographystyle{abbrv-doi}
%\bibliographystyle{abbrv-doi-narrow}
%\bibliographystyle{abbrv-doi-hyperref}
%\bibliographystyle{abbrv-doi-hyperref-narrow}

% \bibliography{refs}

\end{document}